# Effect of atmospheric environment on the attenuation coefficient of light in water

Juan Liu[1], Jiulin Shi[1], Yijun Tang[1], Kaixing Zhu[1], Yuan Ge[1], Xuegang Chen[1], Xingdao He[1], Dahe Liu[1,2]*,

1 Key Laboratory of Nondestructive Test (Ministry of Education), Nanchang Hangkong University, Nachang 330063, China.

2 Applied Optics Beijing Area Major Laboratory, Department of Physics, Beijing Normal University, Beijing 100875, China.

* Corresponding author: Dahe Liu dhliu@bnu.edu.cn

**ABSTRACT**

The attenuation coefficient of 532 nm light in water under different atmospheric conditions was investigated. Measurements were made over a two-year period at the same location and show that the attenuation coefficient is significantly influenced by the atmospheric environment. It is lowest when the atmospheric pressure is high and temperature is low, and is highest when the atmospheric pressure is low and temperature is high. The maximum attenuation coefficient of pure water in these studies was about three times the minimum value. The mechanism of the phenomena is discussed. These results are also important in underwater acoustics.

## 1. Introduction

Optical techniques are very important in oceanographic research involving phenomena such as: Raman scattering [1], Brillouin scattering [2-4], stimulated Brillouin scattering (SBS) [5,6], laser induced breakdown spectroscopy [7], fluorescence spectroscopy including laser induced fluorescence spectroscopy [8,9], bacterial photosynthesis [10], photoheterotrophic bacteria [11], and so forth. In all these research areas, the attenuation of light in the water is a key factor influencing the measurements. Consequently, the attenuation (which includes both absorption and scattering) of light in water has attracted considerable attention. Fry *et. al.* have measured widely accepted data for the absorption of light in pure water [12,13]. Liu *et al.* investigated the attenuation coefficient of light in water [14] and other properties of light in water [15,16] related to SBS. And, the attenuation coefficient of light in deionized water has been measured using a split-pulse laser method[17]. But, the influence of the atmospheric environment on the attenuation coefficient of light in water has not been a subject of much investigation. For example, what relationships are there between the attenuation coefficient in the ocean and the air pressure, temperature, and humidity above the ocean surface. In our lidar studies for remote sensing of the ocean, we found that the observation depth varied significantly with the atmosphere environment. For example, on a sunny winter day, the observation depth for our Brillouin lidar [18] was more than three times the observation depth on a cloudy summer day. It seems clear that the atmospheric environment has a major influence on the propagation of light in water.

Air bubbles in water are a major factor influencing acoustics and light propagation in waterAcoustic bubbles have been studied widely [19-21]; and, many studies have investigated the effects on optical properties due to bubbles that are injected by waves, even breaking waves [22-24]. In contrast, the effect of atmospheric parameters on optical properties in the water has not been considered. Our work shows that changes in the atmosphere environment can, in fact, also induce bubbles in the water that impact optical propagation. The bubbles are induced because atmospheric changes vary the solubility of air in water. Thus, the attenuation of light in water is directly related to the solubility of air in water. Although the solubility of air and different gases in water and seawater has been investigated for several decades, apparently no one has noted the relationship between the solubility of air in water and the attenuation coefficient of light in water. The solubility of air in water is related to the pressure, temperature and humidity of the air [24-28]. In our recent work, these factors played an important role in the attenuation coefficient of light in water; they are investigated in this paper.

**2. Experiment and Results**

The experimental measurements were made using distilled water in an open environment at a fixed location in Nanchang City, China. The water container is made of glass with a wall thickness of 1.5 mm; its dimensions are 200 cm long, 50 cm wide and 50 cm high. The volume of the distilled water sample was $200 \times 50 \times 30$ cm$^3$. To keep the water sample fresh, the container was covered by air ventilation cloth, and the water sample was replaced with fresh distilled water every 24 hours. The container was put under a shade to avoid direct sunlight. Air pressure, temperature and humidity were monitored in real time; while the water temperature was also monitored by inserting a thermometer to a depth of 5 cm below the water surface. The lowest and highest temperatures at Nachang city are about 3℃ and 43℃, respectively. The corresponding the water temperatures ranged from 5℃ to 35℃. When atmospheric conditions had changed, the attenuation coefficient for 532 nm light in the distilled water was measured using the method reported in Ref. [14]. Data were accumulated over a period of two years. Twenty measurements at a specific atmosphere condition would last for about 40 minutes. Generally, in during a 40 minute time period, the atmosphere environment did not change significantly. However, if atmosphere parameters did change during a measurement, the measured results were added to the data group for the other atmosphere condition.

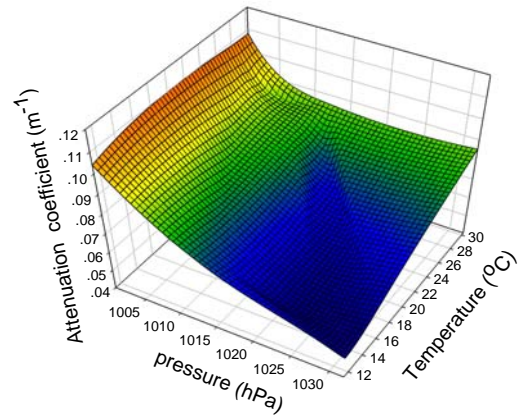

(a) at the humidity of 39%

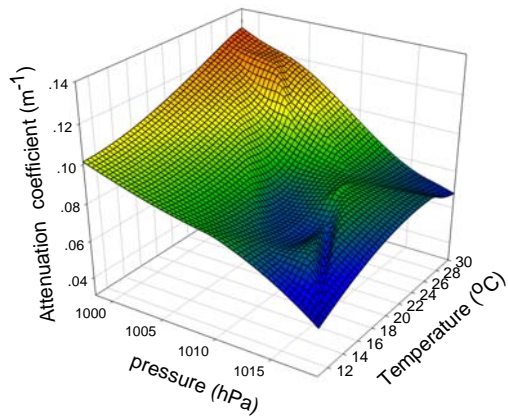

(b) at the humidity of 66%

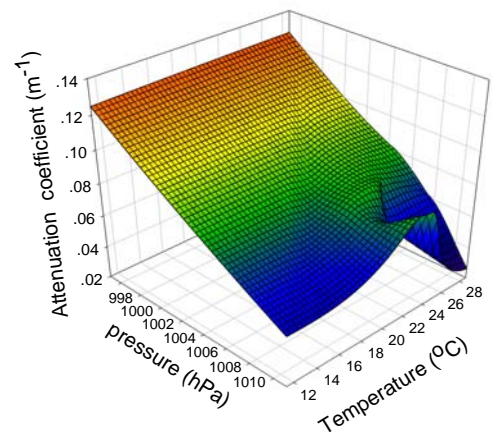

(c) humidity of 83%

Fig. 1  3-D plot of the measured attenuation coefficient of light in water vs. air pressure and water temperature at different humidities.

Figure 1 shows 3-D plots of the data for attenuation coefficient vs. air pressure and water temperature at four different values for the humidity of the air. The value for the humidity shown in the figure is the average over a 3% range in humidity. Although the curved surfaces for different values of the humidity are not exactly the same, they have similar features, *i.e.* the value of the attenuation coefficient tends to decrease with increasing air pressure, and increase with increasing air temperature. The highest attenuations tend to occur at high water temperature and low air pressure.

To more clearly show the influence of humidity, Fig. 2 shows 3-D plots of the data for attenuation coefficient vs. air pressure and the humidity at two different water temperatures, and Fig.3 shows the data for attenuation coefficient vs. humidity of air and water temperature at two different air pressures. At high air pressure and low water temperature, the humidity has little influence on the attenuation of coefficient of light in water. In contrast, at high temperature, the influence becomes more significant.

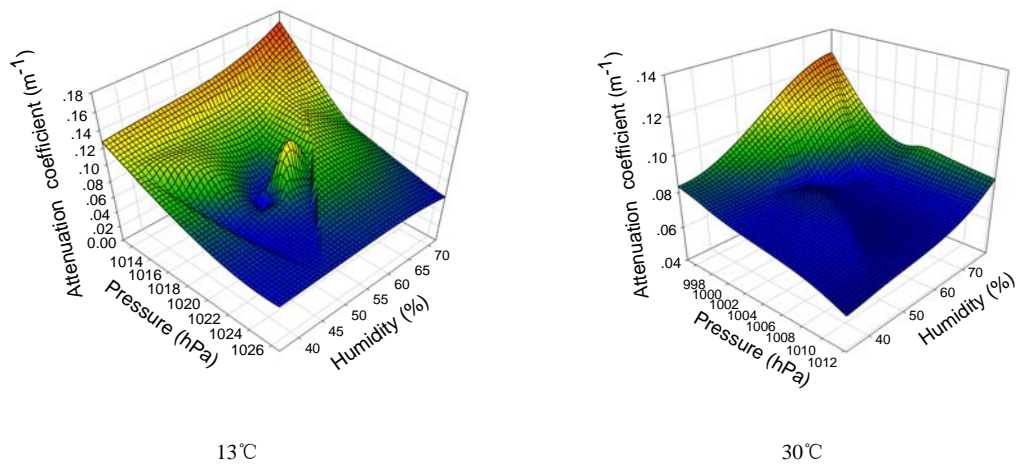

13℃   30℃

Fig. 2   3-D plot of measured attenuation coefficient of light in water vs. the air pressure and the humidity of the air at different water temperature.

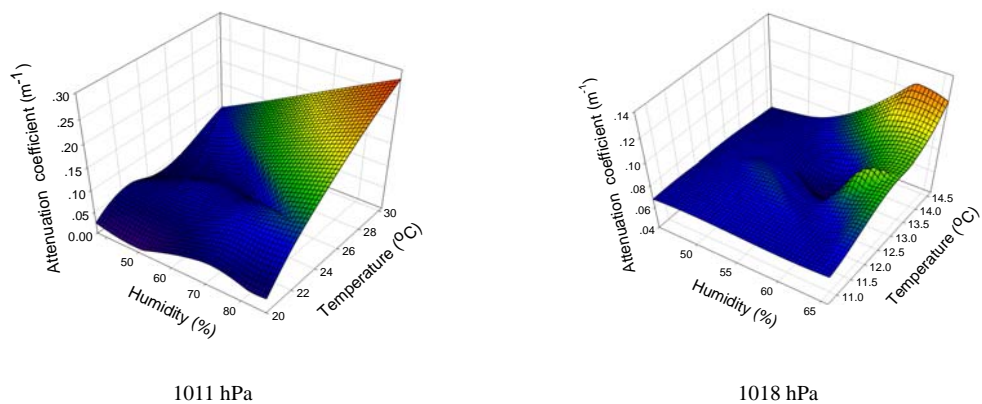

1011 hPa   1018 hPa

Fig. 3   3-D plot of measured attenuation coefficient of light in water vs. water temperature and air humidity of the air at different of air pressures

It should be addressed that, 1) the unit of air pressure in the plots is hPa (=100 Pa ) because the unit of the pressure meter used is hPa; 2) because all measurements were taken under natural conditions, the figures do not have identical axes for the same quantity; 3) all the results shown in the above plots are the average of 20 measured data at the same atmosphere condition. The relative error is defined as the ratio of the root mean square error to the average, the maximum relative error is 17%, the minimum is 1.5%. The error bars are not shown because they tend to be obscure in a 3-D plot.

3. Analysis

Now, consider possible mechanisms leading to the above phenomena. Reported research shows that the solubility of air in water decreases with increasing water temperature, but increases with increasing air pressure [26-28,30]. Fig. 4(a) shows the relationship between solubility and air pressure at different water temperatures; and Fig. 4(b) shows the relationship between the diameter of an air bubble in water and the air pressure [28]. It should be pointed out that Figs 4(a) and (b) cover a wide range of air pressures, but the actual change of the atmosphere pressure is only a small part of it. However, the solubility of air and the bubble size in water still shows considerable change in the small part. Fig. 4(c) shows the solubility of air in water as a function of temperature at atmospheric pressure based on data in table 58 of reference [30]. There is not much data on the formation of bubbles in water as the solubility of air decreases, or on the diameter of air bubbles in water as air pressure changes [28]. Clearly bubble size increases as pressure decreases [31,32], but surface tension plays a major role in the size, *e.g.* the extra pressure inside a 3 micron air bubble is ~1000 hPa, inside a 300 nm bubble it is ~10,000 hPa [33]. In summary, as air pressure increases, the solubility of air in water increases, and bubble diameters decrease. On the other hand, when air pressure decreases, or water temperature increases, the solubility of the air in water decreases, and air dissolved in water will be released via bubbles. An increasing number of bubbles increases the attenuation coefficient; furthermore, if the air pressure is low, the bubble diameter is larger and the attenuation is increased even more.

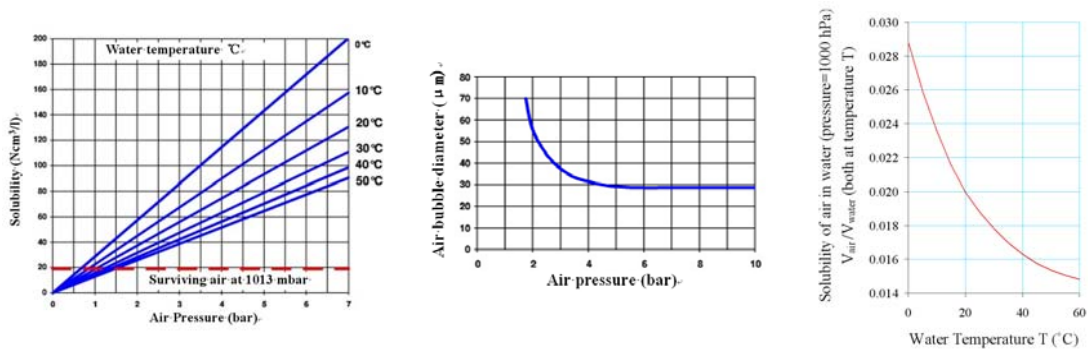

(a) Solubility of air vs. air pressure at different temperatures.
(b) Air bubble diameter vs. air pressure
(c) Solubility of air in water vs. temperature at 1000 hPa

Fig. 4  Air solubility and bubble diameter in water vs. air pressure. The data in (c) is from Table 58 in Ref. [30].

It should be mentioned that the influence of wind [29] was not considered because the water tank is not large enough for the wind to induce waves in it.

## 4. Discussions

Data for the measured attenuation coefficient at different air pressures and water temperatures for three typical values of humidity are given in the following tables. These data are the average of many measurements over the two years of data collection.

Table 1  Data for the attenuation coefficient of light in distilled water at 55% humidity.
Attenuation coefficient (m$^{-1}$)

| Air Pressure (hPa) <br> water Temperature (℃) | 903 | 1000 | 1005 | 1010 | 1015 | 1020 | 1025 | 1030 |
|---|---|---|---|---|---|---|---|---|
| 10 | | | | 0.0716 | 0.0634 | 0.0619 | 0.0578 | 0.0523 |
| 13 | | | | 0.0787 | 0.0697 | 0.0645 | 0.0601 | 0.0542 |
| 16 | | | | 0.0859 | 0.0762 | 0.0700 | 0.0632 | 0.0608 |
| 20 | | | 0.1001 | 0.0903 | 0.0771 | 0.0723 | | |
| 25 | | 0.1127 | 0.1089 | 0.0983 | 0.0858 | 0.0769 | | |
| 30 | 0.1264 | 0.1182 | 0.1138 | 0.1072 | 0.0978 | | | |
| 35 | 0.1290 | 0.1229 | 0.1157 | 0.1101 | 0.0986 | | | |

Table 2  Data for the attenuation coefficient of light in distilled water at 65% humidity.
Attenuation coefficient (m$^{-1}$)

| Air Pressure (hPa) <br> water Temperature (℃) | 903 | 1000 | 1005 | 1010 | 1015 | 1020 | 1025 | 1030 |
|---|---|---|---|---|---|---|---|---|
| 10 | | | | 0.0894 | 0.0863 | 0.0792 | 0.0745 | 0.0639 |
| 13 | | | | 0.0979 | 0.0923 | 0.0834 | 0.0756 | 0.0701 |
| 16 | | | | 0.1002 | 0.0954 | 0.0903 | 0.0828 | 0.0774 |
| 20 | | | 0.1083 | 0.1037 | 0.0958 | 0.0938 | | |
| 25 | | 0.1146 | 0.1112 | 0.1089 | 0.0991 | 0.0969 | | |
| 30 | 0.1272 | 0.1214 | 0.1143 | 0.105 | 0.0998 | | | |
| 31 | 0.1298 | 0.1263 | 0.1192 | 0.1092 | 0.1021 | | | |

Table 3  Data for the attenuation coefficient of light in distilled water at 85% humidity.
Attenuation coefficient (m$^{-1}$)

| Air Pressure (hpa) <br> water Temperature (℃) | 995 | 998 | 1001 | 1004 | 1006 | 1008 | 1010 | 1012 | 1014 | 1016 |
|---|---|---|---|---|---|---|---|---|---|---|
| 10 | | | | | | 0.0948 | 0.0931 | 0.0920 | 0.0908 | 0.0891 |
| 13 | | | | | | 0.1004 | 0.0996 | 0.0989 | 0.0974 | 0.0967 |
| 16 | | | | | 0.1055 | 0.1022 | 0.1015 | 0.1001 | 0.996 | 0.984 |
| 19 | | | | 0.1123 | 0.1109 | 0.1097 | 0.1068 | 0.1054 | 0.1033 | 0.1010 |
| 22 | | | 0.1213 | 0.1200 | 0.1187 | 0.1151 | 0.1137 | 0.1125 | 0.1118 | 0.1100 |
| 25 | | 0.1298 | 0.1285 | 0.1274 | 0.1260 | 0.1244 | 0.1231 | 0.1211 | 0.1200 | |
| 28 | 0.1369 | 0.1355 | 0.1334 | 0.1319 | 0.1301 | 0.1271 | 0.1269 | 0.1234 | | |
| 32 | 0.1462 | 0.1427 | 0.1374 | 0.1305 | 0.1276 | 0.1228 | 0.1157 | | | |

In the three tables, the maximum value of attenuation is 0.1462 m$^{-1}$ at an air pressure of 995 hPa,

water temperature of 32℃ and humidity of 85%; the minimum value is 0.0523 m$^{-1}$ at an air pressure of 1030 hPa, water temperature of 10℃ and humidity of 55%. The maximum is almost the three times of the minimum.

The attenuation coefficient $\gamma$ of light in water as a function of the distance z is given by

$$I(z) = I_0 \exp(-\gamma z) \quad (1)$$

where $I_0$ is the light intensity at z=0. The attenuation length of light in water is defined by $L_\gamma = 1/\gamma$. For distilled water, the maximum $L_\gamma$ corresponding to 0.0523 m$^{-1}$ in Table 1 is 19.12 m. The minimum $L_\gamma$ corresponding to 0.1462 m$^{-1}$ in Table 3 is 6.84 m. The maximum $L_\gamma$ is about three times of the minimum $L_\gamma$. Using our Brillouin lidar [18] the actual observation depth can reach 9 attenuation lengths, a considerable range for lidar. But, even for such a lidar, the actual observation depth has a strong dependence on the atmospheric environment.

Finally, the present investigation is also of significance to underwater acoustics, since air bubbles in water have an important influence on sound waves in water,

**5. Conclusion**

In conclusion, the atmosphere environment has a significant influence on the attenuation coefficient of light in water. Low air pressure and/or high water temperature will significantly increase the attenuation coefficient.

**Acknowledgements**

The authors would like to thank National Advanced Technology Program of China (Grant No. 2009AA09Z101) and National Natural Science Foundation of China (Grant Nos. 41206084，41266001 and 10904003) for the financial support. One of the authors received support from the Robert A. Welch Foundation (grant A-1218).

**References:**

1. D. A. Leonard, B. Caputo and F. E. Hoge, "Remote sensing of subsurface water temperature by Raman scattering", Appl. Opt. 18, 1732-1745 (1979).

2. J. L. Guagliardo and H. L. Dufilho, "Range Resolved Brillouin Scattering a Using Pulsed Laser", Rev. Sci. Instrum., 51, 79-81, (1980).

3. J. G. Hirschberg, J. D. Byrne, A. W. Wouters, and G. C. Boynton, "Speed of sound and temperature in the ocean by Brillouin scattering," Appl. Opt. 23, 2624–2628 (1984).

4. G. D. Hickman, J. M. Harding, M. C. Garnes, A. Pressman, G. W. Kattwar, and E. S. Fry, "Aircraft laser sensing of sound velocity in water: Brillouin scattering," Remote Sens. Environ. 36, 165–178 (1991).

5. Jinwei Shi, Guixin Li, Wenping Gong, Jianhui Bai, Yi Huang, Yinan Liu, Shujing Li, Dahe Liu, "A lidar system based on stimulated Brillouin scattering", Appl. Phys. B 86, 177–179 (2007).

6. Min Ouyang, Jinwei Shi, Luhua Zhao, Xudong Chen, Hongmei Jing, Dahe Liu, "Real time measurement of attenuation coefficient of water in open ocean based on stimulated Brillouin scattering", Appl. Phy. B, Vol.91, 381-385 (2008).

7. A. De Giacomo, M. Dell'Aglio, O. De Pascale, M. Capitelli, "From single pulse to double pulse ns-Laser Induced Breakdown Spectroscopy under water: Elemental analysis of aqueous solutions and submerged solid samples", Spectrochimica Part B, 62, 721–738 (2007).

8. Y. S. Cheng, E. B. Barr, B. J. Fan, P. J. Hargis, D. J, Rader, T. J. O'Hern, J. R. Torczynski, G. C. Tisone, B. L. Preppernau, S. A. Young, and R. J. Radloff, "Detection of biaoaerosols using multiwavelength UV fluorescence spectroscopy", Aerosol Sci. Tech., 31, 409-421 (1999).

9. L. M. Mayer, L. L.Schick, T. C. Loder, "Dissolved protein fluorescence in two main estuaries", Marine Chemistry, 64, 171-179 (1999).

10. Z. S. Kolber, C.L. VanDover, R. A. Niederman, and P. G. Falkowski, "Bacterial photosynthesis in surface warter of the open ocean", Nature, 407, 177-179 (2000).

11. Z. S. Kolber, F. G. Plumley, A. S. Lang, J. T. Beatty, R. E. Blankenship, C. L. VanDover, C. Vetriani, M. Koblizek, C. Rathgeber, P. G. Falkowski1, "Contribution of Aerobic Photoheterotrophic Bacteria to the Carbon Cycle in the Ocean", Science, 292, 2492-2495 (2001).

12. Frank M. Sogandares and Edward S. Fry, "Absorption spectrum (340-640 nm) of pure water. I. Photothermal measurements", Appl. Opt., 36, 8699-8700 (1997).

13. Robin M. Pope and Edward S. Fry, "Absorption spectrum (380-700 nm) of pure water". II. Integrading cavity measurements", Appl. Opt., 36, 8710-8723 (1997).

14. Jianhui Bai, Juan Liu, Yi Huang, Yinan Liu, Lu Sun, Dahe Liu, and E. S. Fry, "Investigations of the attenuation coefficient of a narrow-bandwidth pulsed laser beam in water", Appl. Opt., 46, 6804-5808 (2007).

15. Lei Zhang, Dong Zhang, Zhuo Yang, Jinwei Shi, Dahe Liu, Wenping Gong, and Edward S. Fry, "Experimental investigation on line width compression of stimulated Brillouin scattering in water", Appl. Phys. Lett., 98, 221106 (2011).

16. Jiulin Shi, Yijun Tang, Hongjun Wei, Lei Zhang, Dong Zhang, Jinwei Shi, Wenping Gong, Xingdao He, Kecheng Yang, and Dahe Liu, "Temperature dependence of threshold and gain coefficient of stimulated Brillouin scattering in water", Appl. Phys. B, 108, 717–720 (2012).

17. M. R. Querry, P. G. Cary, and R. C. Waring, "Split-pulse laser method for measuring attenuation coefficients of transparent liquids: application to deionized filtered water in the visible region," Appl. Opt. 17, 3587–3592 (1978).

18. Jinwei Shi, Xudong Chen, Min Ouyang, Juan Liu, Dahe Liu, "Amplification of stimulated Brillouin scattering of two collinear pulsed laser beams with orthogonal polarization", Appl. Opt., Vol.48, 3233-3237 (2009).

19. T. G. Leighton, THE ACOUSTIC BUBBLES, Academic Press, 1994, San Diego.

20. H. Czerski, "An Inversion of Acoustical Attenuation Measurements to Deduce Bubble Populations", J. Atmos. Oceanic Technol, 29, 1139-1148 (2012).

21. Powell Edward, "A Survy of Scattering, Attenuation, and Size Spectra Studies of Bubble Layers and Plumes Beneath the Air-Sea Interface", NRL Memorandum Report 6823, August 1991. *www.dtic.mil/dtic/tr/fulltext/u2/a240262/pdf*

22. Xiangdong Zhang, Marlon Lewis, and Bruce Johnson, "Influence of bubbles on scattering of light in the ocean", Appl. Opt., 37, 6225-6536 (1998).

23. Eric J. Terrill, W. Kendall Melville, and Dariusz Stramski, "Bubble entrainment by breaking waves and their influence on optical scattering in upper ocean", J. Geophys. Res., 106(C8), 16815-16823 (2001).

24. Michael Twardowski, Xiaodong Zhang, Svein Vagle, James Sullivan, Scott Freeman, Hele Czerski, Yu You, Lei Bi, and George Kattawar, "The optical volume scattering function in a surf zone inverted to derive sediment and bubble particle subpopulations", J. Geophys. Res., 117, C00H17 (2012)

25. James H. Carpenter, "NEW MEASUREMENTS OF OXYGEN SOLUBILITY IN PURE AND NATURAL WATER", Limnology and Oceanography, 11, 264-277 (1966).

26. A. Schumpe, "The estimation of gas solubilities in salt solutions", Chemical Engineering Science,

48, 153-158 (1993).

27. EMMERICH WILHELM, RUBlN BATTINO, and ROBERT J. WILCOCK, "Low-Pressure Solubility of Gases in Liquid Water", Chem. Rev., 77, 219-162 (1977).

28. Website: *http://wenku.baidu.com/view/6af928c54028915f804dc2ce.html*

29. Rik Wanninkhof, "Relationship Between Wind Speed and Gas Exchange Over the Ocean", J. Geophys. Res., 97, 7373-7382 (1992).

30. Rubin Battino, Timothy R. Rettich, and Toshihiro Tominaga, "The Solubility of Nitrogen and Air in Liquids", J. Phys. Chem. Ref. Data, **13**, 563-600 (1984).

31. Chris Garrett, Ming Li, David Farmer, "The connection between bubble size spectra and energy dissipation rates in the upper ocean", Journal of Physical Oceanography **30**, 2163-2171 (2000).

32. D. K. Woolf, Bubbles, "Encyclopedia of Ocean Sciences", Eds J.H. Steele, S.A. Thorpe and K.K. Turekian, Academic Press, pp. 352-357 (2001).

33. Pierre-Gilles de Gennes, Francoise Brochard-Wyart, David Quere, "Capillarity and Wetting Phenomena: drops, bubbles, pearls, waves", Springer, New York, pp. 291 (2004).